\documentstyle[sprocl,bbox]{article}

\def\NPA{{\em Nucl. Phys.} A }

\def\PLB{{\em Phys. Lett.}  B }
\def\PRL{{\em Phys. Rev. Lett.} }
\def\PRC{{\em Phys. Rev.} C }
\def\PRD{{\em Phys. Rev.} D }
\def\ZPC{{\em Z. Phys.} C }
\def\half{{\textstyle{1\over 2}}}
\def\be{\begin{equation}}
\def\ee{\end{equation}}
\def\bea{\begin{eqnarray}}
\def\eea{\end{eqnarray}}

\begin{document}

\title{PARTICLE INTERFEROMETRY:\\
       NEW THEORETICAL RESULTS}

\author{Ulrich Heinz}

\address{Institut f\"ur Theoretische Physik, Universit\"at
  Regensburg,\\
  D-93040 Regensburg, Germany}


\maketitle\abstracts{
 By measuring hadronic single-particle spectra and two-particle 
 correlations in heavy-ion collisions, the size and dynamical state of 
 the collision fireball at freeze-out can be reconstructed. I discuss 
 the relevant theoretical methods and their limitations. By applying 
 the formalism to recent pion correlation data from Pb+Pb collisions at 
 CERN we demonstrate that the collision zone has undergone strong 
 transverse growth before freeze-out (by a factor 2-3 in each 
 direction), and that it expands both longitudinally and 
 transversally. From the thermal and flow energy density 
 at freeze-out the energy density at the onset of transverse
 expansion can be estimated from conservation laws. It comfortably 
 exceeds the critical value for the transition to color deconfined 
 matter.  
}

\section{Introduction}
\label{sec1}

In the last few years a large body of evidence has been accumulated 
that the hot and dense collision region in ultrarelativistic heavy ion 
collisions thermalizes and shows collective dynamical behaviour. This 
evidence is based on a comprehensive analysis of the hadronic single 
particle spectra. It was shown that all available data on hadron 
production in heavy ion collisions at the AGS and the SPS can be 
understood within a simple model which assumes locally thermalized 
momentum distributions at freeze-out, superimposed by collective 
hydrodynamical expansion in both the longitudinal and transverse 
directions.\cite{LH89,Stach94} The collective dynamical behaviour in 
the transverse direction is reflected by a characteristic dependence 
of the inverse slope parameters of the $m_\perp$-spectra (``effective 
temperatures'') at small $m_\perp$ on the hadron masses.\cite{LH89} 
New data from the Au+Au and Pb+Pb systems~\cite{QM96} support 
this picture and show that the transverse collective dynamics is much 
more strongly exhibited in larger collision systems than in the 
smaller ones from the first rounds of experiments. The amount of 
transverse flow also appears to increase monotonically with collision 
energy from GSI/SIS to AGS energies, but may show signs of saturation 
at the even higher SPS energy.\cite{QM96}  

The extraction of flow velocities and thermal freeze-out temperatures 
from the measured single particle spectra relies heavily on model 
assumptions.\cite{LH89} There have been alternative suggestions to 
explain the observed features of the hadron spectra without invoking 
hydro-flow.\cite{L96,EHX96} The single-particle spectra are ambiguous 
because they contain no direct information on the space-time structure 
and the space-momentum correlations induced by collective flow. In 
terms of the phase-space density at freeze-out (``emission function'') 
$S(x,p)$ the single-particle spectrum is given by $E\, dN/d^3p = \int 
d^4x\, S(x,p)$; the space-time information in $S$ is completely washed 
out by integration. Thus, on the single-particle level, comprehensive 
model studies are required to show that a simple hydrodynamical model 
with only a few thermodynamic and collective parameters can fit all 
the data, and additional consistency checks are needed to show that 
the extracted fit parameter values lead to an internally consistent 
theoretical picture. The published literature abounds with examples 
demonstrating that without such consistency checks the theoretical 
ambiguity of the single particle spectra is nearly infinite.  

This is the point where Bose-Einstein correlations between the momenta 
of identical particle pairs provide crucial new input. They give 
direct access to the space-time structure of the source {\it and}
its collective dynamics. In spite of some remaining model 
dependence the set of possible model sources can thus be reduced 
dramatically. The two-particle correlation function $C(q,K)$ is 
usually well approximated by a Gaussian in the relative momentum $q$ 
whose width parameters are called ``HBT (Hanbury~Brown-Twiss) radii''.  
It was recently shown~\cite{HB95,CSH95,CNH95} that these radius 
parameters measure certain combinations of the second central 
space-time moments of the source. In general they mix the spatial and 
temporal structure of the source in a nontrivial way,\cite{CSH95} and 
the remaining model dependence enters when trying to unfold these 
aspects.  

Collective dynamics of the source leads to a dependence of the HBT 
radii on the pair momentum $K$; this has been known for many
years,\cite{P84,MS88} but was recently quantitatively reanalyzed,
both analytically~\cite{CSH95,CNH95,AS95,CL96} and
numerically.\cite{WSH96,HTWW96} The velocity gradients associated
with collective expansion lead to a dynamical decoupling of different 
source regions in the correlation function, and the HBT radii measure
the size of the resulting ``space-time regions of homogeneity'' of the 
source~\cite{MS88,AS95} around the point of maximum emissivity for 
particles with the measured momentum $K$. The velocity gradients are 
smeared out by a thermal smearing factor arising from the random 
motion of the emitters around the fluid velocity.\cite{CSH95} Due to 
the exponential decrease of the Maxwell-Boltzmann distribution, this 
smearing factor shrinks with increasing transverse momentum $K_\perp$ 
of the pair, which is the basic reason for the $K_\perp$-dependence of 
the HBT radii. 

Unfortunately, other gradients in the source (for example spatial and 
temporal temperature gradients) can also generate a $K$-dependence of 
the HBT radii.\cite{CSH95,CL96} Furthermore, the pion spectra in 
particular are affected by resonance decay contributions, but only at 
small $K_\perp$. This may also affect the HBT radii in a 
$K_\perp$-dependent way.\cite{Schlei,WH96} The isolation of 
collective flow, in particular transverse flow, from the 
$K_\perp$-dependence of the HBT radii thus requires a careful study of 
these different effects.  

Our group studied this $K$-dependence of the HBT radii within a simple 
analytical model for a finite thermalized source which expands both 
longitudinally and transversally. For presentation I use the 
Yano-Koonin-Podgoretskii (YKP) parametrization of the correlator 
which, for sources with dominant longitudinal expansion, provides an 
optimal separation of the spatial and temporal aspects of the
source.\cite{CNH95,HTWW96} The YKP radius parameters are independent
of the longitudinal velocity of the observer frame. Furthermore, in all 
thermal models without transverse collective flow, they show perfect 
$M_\perp$-scaling (in the absence of resonance decay contributions). 
Only the transverse gradients induced by a non-zero transverse flow 
can break this $M_\perp$-scaling, causing an explicit dependence on 
the particle rest mass. This allows for a rather model-independent 
identification of transverse flow from accurate measurements of the 
YKP correlation radii for pions and kaons. High-quality data should 
also allow to control the effects from resonance decays.  

Due to space limitations, I will be selective with equations,
figures and references. A comprehensive and didactical discussion 
of the formalism and a more extensive selection of numerical examples 
can be found in the lecture notes~\cite{He96} to which I refer the
reader for more details.

\section{Spectra and emission function}
\label{sec2}
\subsection{Single-particle spectra and two-particle correlations}
\label{sec2.1}

The covariant single- and two-particle distributions are defined by
 \begin{eqnarray}
   P_1(\bbox{p}) 
  & = & E\, \frac{dN}{d^3p} 
        = E \, \langle\hat{a}^+_{\bbox{p}} \hat{a}_{\bbox{p}}\rangle \, ,
 \label{1} \\
   P_2(\bbox{p}_a,\bbox{p}_b) 
  & = & E_a\, E_b\, \frac{dN}{d^3p_a d^3p_b}
        = E_a \, E_b\, 
          \langle\hat{a}^+_{\bbox{p}_a} \hat{a}^+_{\bbox{p}_b}
                 \hat{a}_{\bbox{p}_b} \hat{a}_{\bbox{p}_a} \rangle \, ,
 \label{2}
 \end{eqnarray}
where $\hat{a}^+_{\bbox{p}}$ ($\hat{a}_{\bbox{p}}$) creates (destroys) a
particle with momentum $\bbox{p}$. The angular brackets denote an 
ensemble average,
 \begin{equation}
 \label{aver}
    \langle \hat O \rangle = {\rm tr}\, (\hat \rho \hat O)\, ,
 \end{equation} 
where $\hat \rho$ is the density operator associated with the 
ensemble. (When talking about an ensemble we may think of either a 
single large, thermalized source, or a large number of similar, but 
not necessarily thermalized collision events.) The single-particle 
spectrum is normalized to the average number of particles, $\langle N 
\rangle$, per collision, 
 \begin{equation}
 \label{norm1}
   \int {d^3p \over E}\, P_1(\bbox{p}) = \langle N \rangle \, ,
 \end{equation}
while the two-particle distribution is normalized to the number 
of particles in pairs, $\langle N (N-1) \rangle$, per event:
 \begin{equation}
 \label{norm2}
   \int {d^3p_a \over E_a}\,{d^3p_b \over E_b}\, P_2(\bbox{p}_a,\bbox{p}_b) 
   = \langle N (N-1) \rangle \, .
 \end{equation}
The two-particle correlation function is defined as 
 \begin{equation}
 \label{3}
   C(\bbox{p}_a,\bbox{p}_b)
   = \frac{P_2(\bbox{p}_a,\bbox{p}_b)}{P_1(\bbox{p}_a)P_1(\bbox{p}_b)} \, .
 \end{equation}
If the two particles are emitted independently and final state 
interactions are neglected one can prove~\cite{He96} a generalized
Wick theorem
 \begin{equation}
 \label{corr}
  C(\bbox{p}_a, \bbox{p}_b) = 1 \pm 
  {\vert \langle \hat a^+_{\bbox{p}_a} \hat a_{\bbox{p}_b} \rangle \vert^2
   \over
   \langle \hat a^+_{\bbox{p}_a} \hat a_{\bbox{p}_a} \rangle 
   \langle \hat a^+_{\bbox{p}_b} \hat a_{\bbox{p}_b} \rangle } \, .
 \end{equation}
Note that the second term is positive definite, i.e. the correlation 
function cannot, for example, oscillate around unity. This is no
longer true if final state interactions are included (see below).

From now on I will assume that the emitted particles are bosons, and 
for convenience I will call them pions, although nearly everything 
below applies equally well to other bosonic particles.

\subsection{Source Wigner function and spectra} 
\label{sec2.2}

In the language of the covariant current formalism~\cite{GKW79}
the source of the emitted pions can be described in terms of classical
currents $J(x)$ which act as classical sources of freely propagating
pions. These currents represent a parametrization of the last
collision from which the free outgoing pion emerges. Very helpful for
the following will be the so-called ``emission function"
$S(x,K)$:\cite{P84,S73} 
 \begin{equation}
   S(x,K) = \int\frac{d^4y}{2(2\pi)^3}\, e^{-iK{\cdot}y}
   \left\langle J^*(x+\half y)J(x-\half y)\right\rangle \, .
 \label{8f}
 \end{equation}
It is the Wigner transform of the density matrix associated with the 
classical source amplitudes $J(x)$. This Wigner density is a quantum 
mechanical object defined in phase-space $(x,K)$; in general it is
real but not positive definite. But, when integrated over $x$ or 
$K$ it yields the classical (positive definite and real) source 
density in momentum or coordinate space, respectively, in exactly the 
same way as a classical phase-space density would behave. Furthermore, 
textbooks on Wigner functions show that their non-positivity is a
genuine quantum effect resulting from the uncertainty relation and are
concentrated at short phase-space distances; when the Wigner function
is averaged over phase-space volumes which are large compared to the
volume $(2\pi\hbar)^3$ of an elementary phase-space cell, the result
is real and positive definite and behaves exactly like a classical
phase-space density.  

The emission function $S(x,K)$ is thus the quantum mechanical analogue
of the classical phase-space distribution which gives the probability
of finding at point $x$ a source which emits free pions with momentum 
$K$. It allows to express the single-particle spectra and two-particle
correlation function via the following fundamental
relations:\cite{P84,S73,CH94} 
 \begin{eqnarray}
  E_p {dN \over d^3p} &=&
  \int d^4x\, S(x,p) \, ,
 \label{spectrum}\\
  C(\bbox{q},\bbox{K}) &=& 1 + 
  {\left\vert \int d^4x\, S(x,K)\, e^{iq{\cdot}x} \right\vert^2
   \over
   \int d^4x\, S(x,K+\half q) \ \int d^4x\, S(x,K-\half q)}\, .
 \label{correlator}
 \end{eqnarray}
For the single-particle spectrum (\ref{spectrum}), the Wigner function
$S(x,p)$ on the r.h.s. must be evaluated on-shell, i.e. at $p^0=E_p
= \sqrt{m^2 + \bbox{p}^2}$. For the correlator (\ref{correlator}) we 
have defined the relative momentum $\bbox{q} = \bbox{p}_a - 
\bbox{p}_b$, $q^0 = E_a-E_b$ between the two particles in the pair, 
and the total momentum of the pair $\bbox{K} = (\bbox{p}_a + 
\bbox{p}_b)/2$, $K^0= (E_a+E_b)/2$. Of course, since the 4-momenta
$p_{a,b}$ of the two measured particles are on-shell, $p^0_i = E_i =
\sqrt{m^2 + \bbox{p}_i^2}$, the 4-momenta $q$ and $K$ are in general 
off-shell. They satisfy the orthogonality relation
 \begin{equation}
 \label{ortho}
   q \cdot K = 0\,.
 \end{equation}
Thus, the Wigner function on the r.h.s. of Eq.~(\ref{correlator})
is {\em not} evaluated at the on-shell point $K^0 = E_K$. This implies
that for the correlator, in principle, we need to know the off-shell 
behaviour of the emission function, i.e. the quantum mechanical 
structure of the source. Obviously, this makes the problem appear 
rather untractable!  

Fortunately, nature is nice to us: the interesting behaviour of the
correlator (its deviation from unity) is concentrated at small 
values of $\vert \bbox{q} \vert$. Expanding $K^0 = (E_a+E_b)/2$ for 
small $q$ one finds 
 \begin{equation}
 \label{Konshell}
   K^0 = E_K \, \left( 1 + {\bbox{q}^2 \over 8 E_K^2} + 
   {\cal O}\left({\bbox{q}^4 \over E_K^4}\right) \right) 
   \approx E_K \, .
 \end{equation}
Since the relevant range of $q$ is given by the inverse size of
the source (more properly: the inverse size of the regions of 
homogeneity in the source -- see below), the validity of this 
approximation is ensured in practice as long as the Compton wavelength 
of the particles is small compared to this ``source size". For the 
case of pion, kaon, or proton interferometry for heavy-ion collisions 
this is true automatically due to the rest mass of the particles: even 
for pions at rest, the Compton wavelength of 1.4 fm is comfortably 
smaller than any typical nuclear source size. This is of enormous 
practical importance because it allows you essentially to replace the 
source Wigner density by a classical phase-space distribution function
for on-shell particles. This provides a necessary theoretical 
foundation for the calculation of HBT correlations from classical 
hydrodynamic or kinetic (e.g. cascade or molecular dynamics) 
simulations of the collision.  

If the single-particle spectrum is an exponential function of the 
energy then it is easy to prove~\cite{CSH95} that one can replace the 
product of single-particle distributions in the denominator of 
(\ref{correlator}) by the square of the single-particle spectrum 
evaluated at the average momentum $K$:
 \begin{equation}
 \label{corrapp}
  C(\bbox{q},\bbox{K}) \approx 1 + 
  \left\vert {\int d^4x\, e^{iq{\cdot}x}\, S(x,K) 
              \over
              \int d^4x\, S(x,K)} 
  \right\vert^2
  \equiv 1 + \left\vert \langle e^{iq{\cdot}x} \rangle \right\vert^2
  \, .
 \end{equation}
The deviations from this approximation are proportional to the 
curvature of the single-particle distribution in logarithmic 
representation.\cite{CSH95} They are small in practice because the 
measured single-particle spectra are usually more or less exponential.
In the second equality of (\ref{corrapp}) we defined $\langle \dots 
\rangle$ as the average taken with the emission function; due to the 
$K$-dependence of $S(x,K)$ this average is a function of $K$.
This notation will be used extensively in the following.

The fundamental relations (\ref{spectrum}) and (\ref{correlator}) 
resp. (\ref{corrapp}) show that {\em both the single-particle spectrum 
and the two-particle correlation function can be expressed as simple 
integrals over the emission function}. The emission function thus is 
the crucial ingredient in the theory of HBT interferometry: if it is 
known, the calculation of one- and two-particle spectra is 
straightforward (even if the evaluation of the integrals may in some 
cases be technically involved); more interestingly, measurements of 
the one- and two-particle spectra provide access to the emission 
function and thus to the space-time structure of the source. This 
latter aspect is, of course, the motivation for exploiting HBT in 
practice. In my talk I will concentrate on the question to what extent
this access to the space-time structure from only momentum-space data
really works, whether it is complete, and (since we will find it is
not and HBT analyses will thus be necessarily model-dependent) what
can be reliably said about the extension and dynamical space-time
structure of the source anyhow, based on a minimal set of intuitive
and highly suggestive model assumptions.  

\subsection{Final state interactions (FSI)}
\label{sec2.3}

Equation (\ref{correlator}) reflects the absence of final state
interactions (free propagation after emission) by the appearance of
the plane wave factor $e^{iq{\cdot}x}$ under the integral of the
exchange term in the two-particle cross section:
 \begin{eqnarray}
   P_2(\bbox{p}_a,\bbox{p}_b) &=& \int d^4x_1\, d^4x_2\, \Bigl[
   S\left(x_1,K{+}{\textstyle{q\over2}}\right)\,
   S\left(x_2,K{-}{\textstyle{q\over2}}\right) 
 \nonumber\\
   &&\qquad\qquad \pm e^{i q\cdot (x_1-x_2)} \, 
     S(x_1,K)\, S(x_2,K ) \Bigr]\, .
 \label{50}
 \end{eqnarray}
In practice particle interferometry is done with charged particle
pairs which suffer a long-range Coulomb final state repulsion on their
way out to the detector. In addition, there may be strong final state
interactions, e.g. in proton-proton interferometry where there is a
strong $s$-wave resonance just above the two-particle threshold. In
this case Eq.~(\ref{50}) must be replaced by~\cite{AHR97}
 \begin{eqnarray}
  P_2(\bbox{p}_a,\bbox{p}_b) &=&
  \int d^4x\,d^4y\,S\left(x+{\textstyle{y\over 2}},p_a\right)\,
       S\left(x-{\textstyle{y\over 2}},p_b\right) 
 \nonumber\\
  && \times \left[ \theta(y^0) 
  \left\vert \phi_{\bbox{q}/2}(\bbox{y}{-}\bbox{v}_b y^0) \right\vert^2
  + \theta(-y^0) 
  \left\vert \phi_{\bbox{q}/2}(\bbox{y}{-}\bbox{v}_a y^0) \right\vert^2
  \right]
 \nonumber\\
  &\pm& \int d^4x\,d^4y\,S\left(x+{\textstyle{y\over 2}},K\right)\,
       S\left(x-{\textstyle{y\over 2}},K\right)\,
 \nonumber\\
 \label{50+6}
  && \qquad \times\ 
     \phi^*_{-\bbox{q}/2}(\bbox{y}{-}\bbox{v} y^0) \, 
     \phi_{\bbox{q}/2}(\bbox{y}{-}\bbox{v} y^0) .
 \end{eqnarray}
Here $\bbox{v}{=}\bbox{K}/E_K,\, \bbox{v}_a{=}\bbox{p}_a/E_K,\,
\bbox{v}_b{=}\bbox{p}_b/E_K$ are (to quadratic accuracy in $q$) the
velocities of the particles with momentum $\bbox{K},\,\bbox{p}_a,\,
\bbox{p}_b$, respectively, and $\phi_{\bbox{q}/2}(\bbox{r})$
is an FSI distorted wave with asymptotic relative momentum
$\bbox{q}/2$, evaluated at the two-particle relative distance
$\bbox{r}$. Upon replacing the latter by plane waves (\ref{50+6}) turns
into (\ref{50}). The FSI distorted waves can be calculated by solving
a non-relativistic Schr\"odinger equation for the relative motion
which includes the FSI potential {\em in the rest system of the pair}
(where $\bbox{K} = \bbox{v} =0$). Eq.~(\ref{50+6}) represents a
non-relativistic Galilei-transformation of the result from the pair
rest frame to the frame in which $\bbox{p}_a$ and $\bbox{p}_b$ are
measured; therefore it is only valid in observer frames in which the
pair moves non-relativistically. In order to evaluate Eq.~(\ref{50+6})
one must therefore first transform the 4-momenta $p_{a,b}$ to such a
frame (best directly into the pair rest frame). The momentum argument
$\bbox{q}$ of the FSI distorted waves $\phi$ is then the difference
between the two spatial momenta {\em in that frame}, and their
space-time argument $\bbox{y} - \bbox{v}_iy^0$ is the relative
distance of the two particles in that frame {\em at the time when the
  second particle is emitted}.\cite{AHR97} Since the latter depends
not only on the time difference $y^0$ between emission points, but
also on the velocity of the first emitted particle, these arguments
depend on the momentum argument of the emission function associated
with the first emitted particle. The two terms $\sim \theta(\pm y^0)$
in the direct term reflect the two possible time orderings between the
emission points.

\subsection{Implementation in event generators}
\label{sec2.4}

Equations~(\ref{50}) and (\ref{50+6}) can be implemented into event
generators, following the procedure given in:\cite{AHR97,zhang97} 

For the {\em direct term} one selects all pairs $(i,j)$ with $p_i=p_a$, 
$p_j=p_b$ within a given numerical accuracy (bin width) which is 
essentially dictated by event statistics. Each pair is multiplied with 
a weight given by the corresponding probability density $\vert 
\phi_{\bbox{q}/2}\vert^2$ of the FSI distorted wave. The latter must be 
evaluated in a frame in which the pair moves non-relativistically, 
best in the pair rest frame where $\bbox{K} = (\bbox{p}_a + \bbox{p}_b)/2 
=0$. (Then $E_K=m$, and the velocities $\bbox{v}, \bbox{v}_{a,b}$
reduce to their usual nonrelativistic definition.) From the space-time
coordinates $x_i,x_j$ of the pair in the event generator frame one
calculates the distance $\bbox{y}^*_{ij}$ between the two particles in
the pair rest frame at the time when the second particle is
produced. One then computes $\vert\phi_{\bbox{q}^*/2}(\bbox{y}^*_{ij})
\vert^2$ and weights the selected pair $(i,j)$ with this 
number. In this expression $\bbox{q}^*$ is the spatial relative
momentum between the two particles in the pair rest frame which must
be computed from $p_a,p_b$ in the event generator frame. In the
absence of FSI, the corresponding weight is simply 1. The complete
direct term is obtained by summing over all such pairs.   

For the {\em exchange term}, the selection of pairs and weights is a
little less intuitive:\cite{zhang97} One selects all pairs $(i,j)$ with 
$p_i=p_j=K$ (i.e. {\em on-shell particles} (!) with $\bbox{p}_i =
\bbox{p}_j = \bbox{K}$ and $E_i=E_j=E_K$), again within the same
numerical accuracy (bin width) as above. From the production coordinates 
$x_i,x_j$ one again computes the spatial distance $\bbox{y}^*_{ij}$ 
between the two particles in the pair at the time of emission of the 
second one, in the pair rest frame $\bbox{K}=0$. This distance is used 
to compute the weight $\phi^*_{-\bbox{q}^*/2}(\bbox{y}^*_{ij})
 \phi^*_{\bbox{q}^*/2}(\bbox{y}^*_{ij})$ for this pair. The value of
$\bbox{q}^*$ here is {\em the same as above in the direct term},
i.e. it is computed from $p_a$ and $p_b$ by transforming into the pair
rest frame, not from $p_i=p_j=K$. Without FSI, the corresponding
weight~\cite{zhang97} is $\cos(\bbox{q}^*\cdot\bbox{y}_{ij}^*)$. The
full exchange term is obtained by summing over all such pairs. Note
that this selection of pairs and weights differs from previously
applied algorithms which were shown~\cite{zhang97} to yield wrong
results for sources with very strong $x$-$p$ correlations.

Finally one must normalize the correlator by the product of single
particle spectra,
 \begin{equation}
 \label{single}
   P_1(\bbox{p}_a)\,P_1(\bbox{p}_b) = 
   \int d^4x\, S(x,p_a)\,\int d^4y\, S(y,p_b)\, .
 \end{equation}
This normalization is best obtained from the pairs selected for the
direct term above by multiplying them with unit weights. 

One may object to the use of event generators for the emission 
function because they fix particle momenta and coordinates 
simultaneously and thus violate the uncertainty principle. One can
generate from an event generator a quantum mechanically consistent
Wigner density $S(x,p)$ by folding the event generator output with
minimum uncertainty wave packets.\cite{zhang97} The corresponding
quantum mechanically consistent algorithm~\cite{zhang97} for computing
single- and two-particle spectra is easily generalized to include FSI
effects, by simply replacing the factors 1 and
$\cos(\bbox{q}^*\cdot\bbox{y}^*_{ij})$ in the direct and exchange  
terms, respectively, by the correct FSI weights as discussed above.  

At this point I will drop the discussion of final state interactions;
the rest of the lecture will deal only with the case of free
particles, assuming (carelessly) that appropriate Coulomb corrections
of the data have already been done by the experimentalists.

\subsection{The mass-shell constraint}
\label{sec2.5}

Expressions (\ref{correlator},\ref{corrapp}) show that the correlation 
function is related to the emission function by a Fourier 
transformation. At first sight this might suggest that one should 
easily be able to reconstruct the emission function from the measured 
correlation function by inverse Fourier transformation, the single 
particle spectrum (\ref{spectrum}) providing the normalization. This 
is, however, not correct. The reason is that, since the correlation 
function is constructed from the on-shell momenta of the measured 
particle pairs, not all four components of the relative momentum $q$ 
occurring on the r.h.s. of (\ref{corrapp}) are independent. They are 
related by the ``mass-shell constraint" (\ref{ortho}) which can, for 
instance, be solved for $q^0$:
 \begin{equation}
 \label{massshell}
   q^0 = \bbox{\beta}\cdot \bbox{q} \qquad {\rm with} \qquad 
   \bbox{\beta} = {\bbox{K}\over K^0} \approx {\bbox{K}\over E_K}\, .
 \end{equation} 
$\bbox{\beta}$ is (approximately) the velocity of the c.m. of the 
particle pair. The Fourier transform in (\ref{corrapp}) is therefore 
not invertible, and the reconstruction of the space-time structure of 
the source from HBT measurements will thus always require additional 
model assumptions. 

It is instructive to insert (\ref{massshell}) into (\ref{corrapp}): 
 \begin{equation}
 \label{corrapp1}
  C(\bbox{q},\bbox{K}) \approx 1 + 
  \left\vert {\int d^4x\, \exp\bigl( 
                          i\bbox{q}{\cdot}(\bbox{x}-\bbox{\beta}\, t) 
                          \bigr) \, S(x,K)
              \over
              \int d^4x\, S(x,K)} 
  \right\vert^2 \, .
 \end{equation}
This shows that the correlator $C(\bbox{q},\bbox{K})$ actually mixes the 
spatial and temporal information on the source in a non-trivial way 
which depends on the pair velocity $\bbox{\beta}$. Only for 
time-independent sources things seem to be simple: the correlator then 
just measures the Fourier transform of the spatial source 
distribution. Closer inspection shows, however, that it does so only 
in the directions {\em perpendicular} to $\bbox{\beta}$ since the time 
integration leads to a $\delta$-function 
$\delta(\bbox{\beta}{\cdot}\bbox{q})$: 
 \begin{equation}
 \label{static}
  \lim_{T\to \infty} \left\vert 
  {\int_{-T}^T dt\, \exp\left(-i\,\bbox{q}{\cdot}\bbox{\beta}\, t\right) 
   \over 
   \int_{-T}^T dt} \right\vert^2 =  
  \lim_{T\to \infty} {2\pi \over T} \,
  \delta(\bbox{q}{\cdot}\bbox{\beta}) \, .
 \end{equation}
This implies that there are no correlations in the direction {\em 
parallel} to the pair velocity $\bbox{\beta}$ (which will be called 
the ``outward" direction below), i.e. $C=1$ for $q_{\rm out}\ne 0$. 
The width of the correlator in this direction vanishes! This should 
puzzle you: wouldn't you have thought that the width of the correlator 
in the ``outward" direction is inversely related to the source size in 
that direction (which is, of course, perfectly finite)? As we will 
see in the next subsection this unexpected behaviour is just another 
consequence of the mixing of the spatial and temporal structure of the 
source in the correlator: The width parameter of the correlator in the 
``outward" direction receives also a contribution from the lifetime of 
the source which in this case diverges, leading to the vanishing width 
of the correlator.  

It is instructive to look at the problem also in the following way:
If one rewrites Eq.~(\ref{corrapp1}) in the pair rest frame where
$\bbox{K}=0$ and hence $q^0=0$, one obtains 
 \begin{equation}
 \label{restframe}
   C(\bbox{q,K})- 1 = \int d^3r\, \cos(\bbox{q\cdot r})\,
   S_{\rm rel}(\bbox{r};K)
 \end{equation}
where
 \begin{equation}
 \label{reldis}
   S_{\rm rel}(\bbox{r};K) = \int d^3R\  
   \bar s_K(\bbox{R} + {\textstyle{1\over 2}} \bbox{r})\ 
   \bar s_K(\bbox{R} - {\textstyle{1\over 2}} \bbox{r})\, ,
 \end{equation}
with
 \begin{equation}
 \label{sbar}
   \bar s_K(\bbox{x}) = \int dt\ s(\bbox{x},t; K) 
   = \int dt\ {S(x,K) \over \int d^4x'\, S(x',K)}\, ,
 \end{equation}
is the {\em time-integrated} normalized relative distance distribution
in the source. The latter can, in principle, be uniquely reconstructed
from the measured correlator~\cite{brown97} by inverting the
cosine-Fourier transform (\ref{restframe}). But since it gives only
the time integral of the relative distance distribution for fixed pair
momentum $\bbox{K}$ in the pair rest frame, no direct information on
the time structure of the source is obtainable! Only by looking at the
result as a function of $\bbox{K}$, which, as I will show, brings out
the collective dynamical features of the source, can one hope to
unfold the time-dependence of the emission function. It is clear that
this will be only possible within the context of specific source
parametrizations. 

\subsection{$K$-dependence of the correlator}
\label{sec2.6}

We have seen that in general the correlator is a function 
of {\em both} $\bbox{q}$ and $\bbox{K}$. Only if the emission function 
factorizes in $x$ and $K$, $S(x,K) = F(x)\,G(K)$, which means that every 
point $x$ in the source emits particles with the same momentum 
spectrum $G(K)$ (no ``$x$-$K$-correlations"), the $K$-dependence in 
$G(K)$ cancels between numerator and denominator of (\ref{corrapp}), 
and the correlator seems to be $K$-independent. However, not even 
this is really true: even after the cancellation of the explicit 
$K$-dependence $G(K)$, there remains an implicit $K$-dependence
via the pair velocity $\bbox{\beta} \approx \bbox{K}/E_K$ 
in the exponent on the r.h.s. of Eq.~(\ref{corrapp1})! Only if both 
conditions, factorization of the emission function in $x$ and $K$ {\em 
and} time-independence of the source, apply simultaneously, the 
correlation function is truely $K$-independent (because then the 
$\bbox{\beta}$-dependence resides only in the $\delta$-function 
(\ref{static})). 

The only practical situation which I know where this occurs and a 
$K$-independent correlation function should thus be expected is in HBT 
interferometry of stars for which the method was invented.\cite{HBT}
It is hard to believe that this complication in the application of the 
original HBT idea to high-energy collisions went nearly unnoticed for 
more than 20 years and was stumbled upon more or less empirically by 
Scott Pratt in his pioneering work on HBT interferometry for heavy-ion 
collisions~\cite{P84} only in 1984!  

If one parametrises it by a Gaussian in $q$ (see below) this means 
that in general the parameters (``HBT radii'') depend on $K$. Typical
sources of $x$-$K$ correlations in the emission function are a
collective expansion of the emitter and/or temperature gradients in 
the particle source: in both cases the momentum spectrum $\sim \exp[-
p{\cdot}u(x)/T(x)]$ of the emitted particles (where $u^\mu(x)$ is the 
4-velocity of the expansion flow) depends on the emission point. In 
the case of collective expansion, the spectra from different emission 
points are Doppler shifted relative to each other. If there are
temperature gradients, e.g. a high temperature in the center and 
cooler matter at the edges, the source will look smaller for 
high-momentum particles (which come mostly from the hot center) than 
for low-momentum ones (which receive larger contributions also from 
the cooler outward regions).

We thus see that collective expansion of the source induces a 
$K$-dependence of the correlation function. But so do temperature 
gradients. The crucial question is: does a careful measurement of the 
correlation function, in particular of its $K$-dependence, permit a 
separation of such effects, i.e. can the collective dynamics of the 
source be quantitatively determined through HBT experiments? We will 
see that this is not an easy task; however, with sufficiently good 
data, it should be possible. In any case, the $K$-dependence of the 
correlator is a decisive feature which puts the HBT game into a 
completely new ball park. Two-particle correlation measurements which
are not able to resolve the $K$-dependence of the HBT parameters are,
in high energy nuclear and particle physics, of very limited use only.
[Unfortunately, this applies to all published HBT data from $pp$ and
$e^+e^-$ collisions. In my opinion, a renewed investigation of
two-particle correlations from $pp$ and $e^+e^-$ collisions, using the
powerful new tool of multidimensional, differential HBT analysis,
should be a high priority project -- as it is, we have practically
nothing with which to compare our heavy-ion results in a meaningful
way.] 

\section{Model-independent discussion of HBT correlation functions}
\label{sec3}
\subsection{The Gaussian approximation}
\label{sec3.1}

The most interesting feature of the two-particle correlation function
is its half-width. Actually, since the relative momentum $\bbox{q} =
\bbox{p}_1 - \bbox{p}_2$ has three Cartesian components, the fall-off
of the correlator for increasing $q$ is not described by a single
half-width, but rather by a (symmetric) 3$\times$3 tensor~\cite{CNH95}
which describes the curvature of the correlation function near
$\bbox{q} = 0$. We will see that in fact nearly all relevant
information that can be extracted from the correlation function
resides in the 6 independent components of this tensor. This in turn
implies that in order to compute the correlation function $C$ it is
sufficient to approximate the source function $S$ by a Gaussian in $x$
which contains only information on its space-time moments up to second
order.  

Let us write the arbitrary emission function $S(x,K)$ in the following
form: 
 \begin{equation}
 \label{7}
   S(x,K) = N(K)\  S(\bar x(K),K)\ 
            e^{ - \half \tilde x^\mu(K)\, B_{\mu\nu}(K)\, \tilde x^\nu(K)}
   + \delta S(x,K) \, ,
 \end{equation} 
where we adjust the parameters $N(K)$, $\bar x^\mu(K)$, and 
$B_{\mu\nu}(K)$ of the Gaussian first term in such a way that the 
correction term $\delta S$ has vanishing zeroth, first and second 
order space-time moments:
 \begin{equation}
 \label{deltaS}
   \int d^4x\, \delta S(x,K) = 
   \int d^4x\, x^\mu\, \delta S(x,K) = 
   \int d^4x\, x^\mu x^\nu\,  \delta S(x,K) = 0\, .
 \end{equation} 
This is achieved by choosing
 \begin{eqnarray}
  N(K) &=& E_K {dN\over d^3 K}\,
           {\det^{1/2} B_{\mu\nu}(K) \over S(\bar x(K),K)}\, ,
 \label{NK}\\
  \bar x^\mu(K) &=& \langle x^\mu \rangle\, , 
 \label{barx}\\
  \left(B^{-1}\right)_{\mu\nu}(K) 
  &=& \langle \tilde x_\mu \tilde x_\nu \rangle 
      \equiv \langle (x -\bar x)_\mu (x- \bar x)_\nu \rangle \, .
 \label{Bmunu}
 \end{eqnarray}
The ($K$-dependent) average over the source function $\langle \dots 
\rangle$ has been defined in Eq.~(\ref{corrapp}). The normalization factor 
(\ref{NK}) ensures that the Gaussian term in (\ref{7}) gives the 
correct single-particle spectrum (\ref{spectrum}); it fixes the 
normalization on-shell, i.e. for $K^0=E_K$, but as we discussed this 
is where we need the emission function also for the computation of the 
correlator. $\bar x(K)$ in (\ref{barx}) is the centre of the emission
function $S(x,K)$ and approximately equal to its ``saddle point",
i.e. the point of highest emissivity for particles with momentum $K$. The 
second equality in (\ref{Bmunu}) defines $\tilde x$ as the space-time 
coordinate relative to the centre of the emission function; only this 
quantity enters the further discussion, since, due to the invariance 
of the momentum spectra under arbitrary translations of the source in 
coordinate space, the absolute position of the emission point is not 
measurable in experiments which determine only particle momenta. Since 
$\bar x(K)$ is not measurable, neither is the normalization
$N(K)$~\cite{HTWW96} as its definition (\ref{NK}) involves the
emission function at $\bar x(K)$. Finally, Eq.~(\ref{Bmunu}) ensures that the 
Gaussian first term in (\ref{7}) correctly reproduces the second
central space-time moments $\langle \tilde x_\mu \tilde x_\nu \rangle$
of the original emission function, in particular its r.m.s. widths in
the various space-time directions.  

Inserting the decomposition (\ref{7}) into Eq.~(\ref{corrapp}) we 
obtain for the correlation function
 \begin{equation}
 \label{corrgauss}
   C(\bbox{q},\bbox{K}) = 1 + \exp\bigl[
   - q^\mu\, q^\nu\, \langle \tilde x_\mu \tilde x_\nu \rangle(\bbox{K}) 
   \bigr]
   + \delta C(\bbox{q},\bbox{K})\, .
 \end{equation}
The Gaussian in $q$ results from the Fourier transform of the Gaussian 
contribution in (\ref{7}); the last term $\delta C$ receives 
contributions from the second term $\delta S$ in (\ref{7}) which 
contains information on the third and higher order space-time moments 
of the emission function, like sharp edges, wiggles, secondary peaks,
or non-Gaussian tails in the source. It is at least of fourth order in
$q$, i.e. the second derivative of the full correlator at $q=0$ is
given {\em exactly} by the Gaussian in (\ref{corrgauss}). Please note
that the exponent of the correlator contains no term linear in $q$;
since the correlator must be symmetric under $\bbox{q} \to - \bbox{q}$
because it does not matter which of the two particles of the pair
receives the label 1 or 2, a linear $q$-dependence could only arise in
the form $\exp(-R \vert \bbox{q} \vert)$. The only type of emission
function yielding such a $q$-dependence of the correlator would be a 
spherically symmetric Lorentzian. Any emission function which at large 
$x$ falls off faster than $1/x^2$ results in the leading Gaussian 
behaviour (\ref{corrgauss}) instead. This settles, in my opinion, the 
old issue whether Gaussian or exponential fits of the correlation 
function should be preferred.  

[In the past it has repeatedly been observed that the correlation data 
appear to be better fit by exponentials than by Gaussians. However, 
as far as I know, this happened always when one tried to fit the 
correlator as a function of the single Lorentz invariant variable 
$Q_{\rm inv}^2 = (q^0)^2 - \bbox{q}^2$. Contemplating the structure
of Eq.~(\ref{corrgauss}) one realizes that such a fit does not make 
sense: the generic structure of the exponent, $-q^\mu q^\nu \langle 
\tilde x_\mu \tilde x_\nu \rangle$, tells us that the term $(q^0)^2$ 
should come with the time variance of the source while the spatial 
components $(q^i)^2$ should come with the spatial variances of the 
source. Since all variances are positive semidefinite by definition, 
it does not make sense to parametrize the correlation function by a 
variable in which $(q^0)^2$ and $\bbox{q}^2$ appear with the opposite 
sign! Such a fit could only work if the time variance and all mixed
variances would vanish identically, and all three spatial variances 
were equal. This is certainly not the general case in nature. The
good exponential fits of the correlation functions from $pp$ and 
$e^+e^-$ collisions are thus, in my mind, purely accidental and an 
empirical curiosity without physical meaning. {\it The variable 
$Q_{\rm inv}$ should {\em not} be used for fitting HBT data.}]  

Please note also that Eq.~(\ref{corrgauss}) has no factor $\half$ in 
the exponent. If the measured correlator is fitted by a
Gaussian as defined in (\ref{corrgauss}), its $q$-width 
can be directly interpreted in terms of the r.m.s. widths of the 
source in coordinate space. Any remaining factors of $\sqrt{2}$, 
$\sqrt{3}$, or $\sqrt{5}$ (which you can sometimes find in the 
literature) are due to reexpressing the r.m.s. width of the source in 
terms of certain other width parameters chosen for the parametrization 
of the source in coordinate space. The confusion connected with such 
factors is easily avoided by always expressing the source 
parametrization directly in terms of r.m.s. widths.  

Eqs.~(\ref{7}) and (\ref{corrgauss}) would, of course, not be useful 
if the contributions from $\delta S$ and $\delta C$ were not somehow 
small enough to be neglected. This requires a numerical investigation.
It was shown~\cite{WSH96} that in typical (and even in some not so
typical) situations $\delta S$ has a negligible influence on the half
width of the correlation function. It contributes only weak, 
essentially unmeasurable structures in $C(\bbox{q},\bbox{K})$ at large
values of $\bbox{q}$. The reader can easily verify this analytically
for an emission function with a sharp box profile; the
results~\cite{CNH95} for the exact correlator and the one resulting
from the Gaussian approximation (\ref{7}) differ by less than 5\% in
the half widths; the exact correlator has, as a function of $q$,
secondary maxima with an amplitude below 5\% of the value of the
correlator at $q=0$. We have checked that similar statements remain
even true for a source with a doughnut structure, i.e. with a hole in
the middle, which was obtained by rotating the superposition of two
1-dimensional Gaussians separated by twice their r.m.s. widths around
their center. The only situation where these statements require
qualification is if the correlator receives contributions from the
decay of long-lived resonances; unfortunately, this is of relevance
for pion interferometry as will be discussed in Sec.~ref{sec6}.

From Eq.~(\ref{corrgauss}) we conclude that the two-particle 
correlation function measures the second central space-time moments
of the emission function. That's it -- finer features of its 
space-time structure (edges, wiggles, holes) cannot be measured with 
two-particle correlations, but require the analysis of three-, four-,
\dots, many-particle correlations.~\cite{HZ97} The variances $\langle
\tilde x_\mu \tilde x_\nu \rangle$ are in general {\em not} identical
with our naive intuitive notion of the ``source radius": unless the
source is stationary and has no $x$-$K$-correlations at all, the
variances depend on the momentum $\bbox{K}$ of the pair and cannot be 
interpreted in terms of simple overall source geometry. Their correct 
physical interpretation~\cite{CSH95,MS88,AS95} is in terms of 
``lengths of homogeneity" which give, for each pair momentum 
$\bbox{K}$, the size of the region around the point of maximal 
emissivity $\bar x(\bbox{K})$ over which the emission function is 
sufficiently homogeneous to contribute to the correlation function. 
Thus HBT measures ``regions of homogeneity" in the source and their 
variation with the momentum of the particle pairs. As we will see, the 
latter is the key to their physical interpretation.  

\subsection{YKP parametrization for the correlator and HBT radius
  parameters}
\label{sec3.2}

A full characterization of the source in terms of its second order 
space-time variances requires knowledge of the 10 parameters $\langle 
\tilde x_\mu \tilde x_\nu \rangle$. These quantities appear in the 
expression (\ref{corrapp}) for the correlation function but this 
expression still uses all four components of the relative momnetum 
$q^\mu$. However, as already noted only three of the four components 
are independent, due to the mass-shell constraint (\ref{massshell}).
Thus only 6 linear combinations of the variances $\langle 
\tilde{x}_\mu \tilde{x}_\nu \rangle(\bbox{K})$ are actually
measurable.\cite{CNH95}

If the source is azimuthally symmetric around the beam axis, this 
counting changes as follows: Even if the source is azimuthally 
symmetric in coordinate space, the emission function $S(x,K)$ in phase 
space is for finite $\bbox{K}$ no longer azimuthally symmetric because 
the transverse components $\bbox{K}_\perp$ of the pair momentum 
distinguish a direction transverse to the beam direction. There 
remains, however, a reflection symmetry with respect to the plane 
spanned by $\bbox{K}$ and the beam axis. If we call the direction 
orthogonal to this plane $y$, all mixed variances which are linear in 
$y$ must vanish due to this reflection symmetry, and the correlator 
must be symmetric under $q_y \to -q_y$. Thus only 7 non-vanishing 
variances $\langle \tilde x_\mu \tilde x_\nu \rangle$ survive in 
general, of which, due to the mass-shell constraint (\ref{massshell}) 
only 4 linear combinations are measurable.  

Before the correlator (\ref{corrgauss}) can be fit to experimental 
data, the redundant components of $q$ must first be eliminated from 
the exponent of the Gaussian via (\ref{massshell}). We use a cartesian
coordinate system with the $z$-axis along the beam direction and the
$x$-axis along $\bbox{K}_\perp$. Then $\bbox{\beta} = (\beta_\perp, 0,
\beta_l)$. We assume an azimuthally symmetric source (impact parameter
$\approx 0$) and eliminate from (\ref{corrgauss}) $q_x$ and $q_y$ in
terms of $q_\perp^2 = q_x^2 + q_y^2$, $q_l$ and $q^0$. This yields the
YKP parametrization:\cite{CNH95,HTWW96}
 \begin{equation}
 \label{18}
   C(\bbox{q,K}) = 1 +  
     \exp\biggl[ - R_\perp^2 q_\perp^2 
                 - R_\parallel^2 \left( q_l^2 - (q^0)^2 \right) 
                 - \left( R_0^2 + R_\parallel^2 \right)
                         \left(q\cdot U\right)^2
                \biggr] .
 \end{equation}
Here $R_\perp$, $R_\parallel$, $R_0$, $U$ are four $K$-dependent
parameter functions. $U(\bbox{K})$ is a 4-velocity with only a
longitudinal spatial component:  
 \begin{equation}
 \label{19}
   U(\bbox{K}) = \gamma(\bbox{K}) \left(1, 0, 0, v(\bbox{K}) \right) ,
   \ \ {\rm with} \ \
   \gamma = {1\over \sqrt{1 - v^2}}\, .
 \end{equation}
Its value depends, of course, on the measurement frame. The ``Yano-Koonin 
velocity'' $v(\bbox{K})$ can be calculated~\cite{HTWW96} in an
arbitrary reference frame from the second central space-time moments
of $S(x,K)$. It is, to a good approximation, the longitudinal velocity
of the fluid element from which most of the particles with momentum
$\bbox{K}$ are emitted.\cite{CNH95,HTWW96} For sources with
boost-invariant longitudinal expansion velocity the YK-rapidity
associated with $v(\bbox{K})$ is linearly related to the pair rapidity
$Y$.\cite{HTWW96}

The other three YKP parameters do not depend on the longitudinal 
velocity of the observer. (This distinguishes the YKP form (\ref{18})
from the Pratt-Bertsch parametrization~\cite{HB95,CSH95,P84} which
results from eliminating $q^0$ in (\ref{corrgauss}).) Their physical
interpretation is easiest in terms of coordinates measured in the
frame where $v(\bbox{K})$ vanishes. There they are given by~\cite{CNH95} 
 \begin{eqnarray}   
   R_\perp^2(\bbox{K}) &=& \langle \tilde y^2 \rangle \, ,
 \label{20a} \\
   R_\parallel^2(\bbox{K}) &=& 
   \left\langle \left( \tilde z - (\beta_l/\beta_\perp) \tilde x
                \right)^2 \right \rangle   
     - (\beta_l/\beta_\perp)^2 \langle \tilde y^2 \rangle 
     \approx \langle \tilde z^2 \rangle \, ,
 \label{20b} \\
   R_0^2(\bbox{K}) &=& 
   \left\langle \left( \tilde t -  \tilde x/\beta_\perp
                \right)^2 \right \rangle 
    -  \langle \tilde y^2 \rangle/\beta_\perp^2 
    \approx \langle \tilde t^2 \rangle .
 \label{20c}
 \end{eqnarray}
$R_\perp$, $R_\parallel$ and $R_0$ thus measure, approximately, the 
($K$-dependent) transverse, longitudinal and temporal regions of 
homogeneity of the source in the local comoving frame of the emitter. 
The approximation in (\ref{20b},\ref{20c}) consists of dropping terms 
which for the model discussed below vanish in the absence of
transverse flow and were found to be small even for finite transverse
flow.\cite{HTWW96} Note that it leads to a complete separation of 
the spatial and temporal aspects of the source. This separation is
spoiled by sources with $\langle \tilde x^2 \rangle \ne \langle \tilde
y^2 \rangle$. For our source this happens for non-zero transverse (in
particular for large) transverse flow $\eta_f$, but for opaque sources
where particle emission is surface dominated~\cite{HV96} this occurs
even without transverse flow.\cite{HV96,TH97}

\section{A model for a finite expanding source}
\label{sec4}

\begin{figure}
\vspace*{11.5cm}
\includegraphics{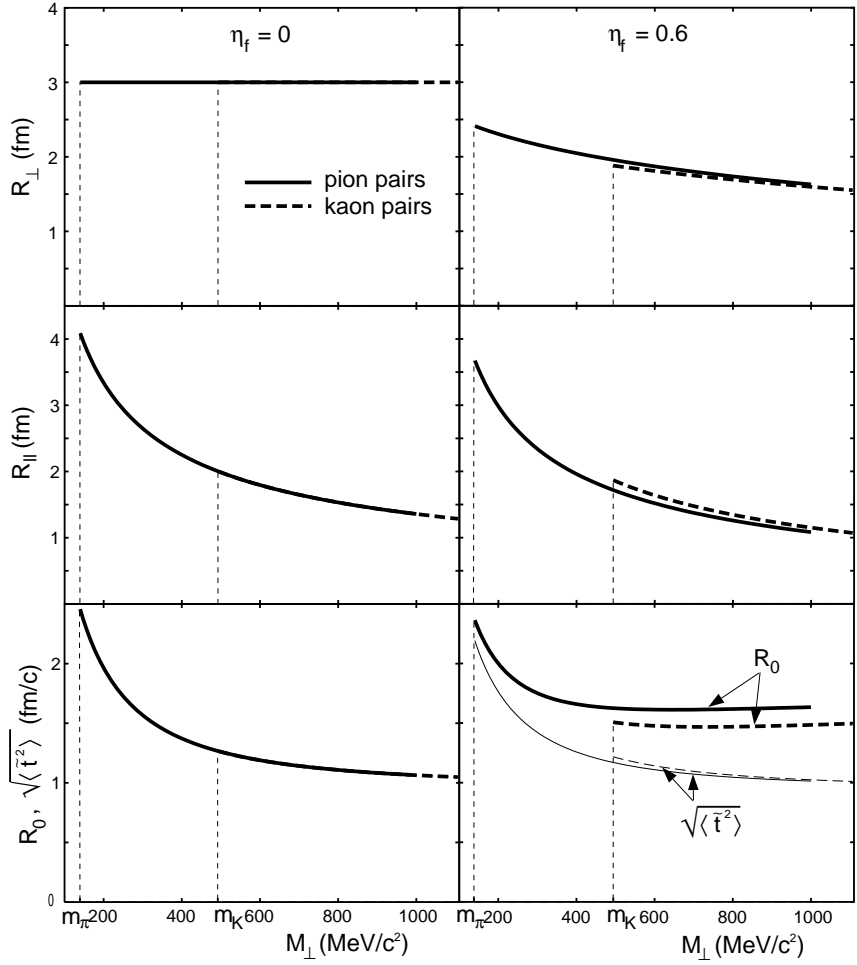}
\caption{The YKP radii $R_\perp$, $R_\parallel$, and $R_0$ (top
  to bottom) for zero transverse flow (left column) and for
  $\eta_f=0.6$ (right column), as functions of $M_\perp$ for pairs at
  $Y_{\rm cm}=0$. Solid (dashed) lines are for pions (kaons). The
  breaking of the $M_\perp$-scaling by transverse flow is obvious in
  the right column. For nonzero transverse flow $R_0$ also does not
  agree exactly with the effective source lifetime $\sqrt{\langle \tilde
  t^2\rangle}$ (lower right panel). Source parameters: $T=140$ MeV,
  $\Delta\eta=1.2$, $R=3$ fm, $\tau_0=3$ fm/$c$, $\Delta\tau=1$ fm/$c$.
  \label{F1}}
\end{figure}

For our quantitative studies we used the following model for an
expanding thermalized source:\cite{CNH95}  
 \begin{equation}
 \label{3.15}
    S(x,K)\! =\! {M_\perp \cosh(\eta{-}Y) \over 8 \pi^4 \Delta \tau}
    \exp\!\! \left[- {K{\cdot}u(x) \over T(x)}
                       - {(\tau-\tau_0)^2 \over 2(\Delta \tau)^2}
                       - {r^2 \over 2 R^2} 
                       - {(\eta- \eta_0)^2 \over 2 (\Delta \eta)^2}
           \right]
 \end{equation}
Here $r^2 = x^2+y^2$, the spacetime rapidity $\eta = {1 \over 2} 
\ln[(t+z)/(t-z)]$, and the longitudinal proper time $\tau= \sqrt{t^2-
z^2}$ parametrize the spacetime coordinates $x^\mu$, with measure 
$d^4x = \tau\, d\tau\, d\eta\, r\, dr\, d\phi$. $Y = {1\over 2} 
\ln[(E_K+K_L)/(E_K-K_L)]$ and $M_\perp = \sqrt{m^2 + K_\perp^2}$ 
parametrize the longitudinal and transverse components of the pair 
momentum $\bbox{K}$. $\sqrt{2} R$ is the transverse geometric
(Gaussian) radius of the source, $\tau_0$ its average freeze-out
proper time, $\Delta \tau$ the mean proper time duration of particle
emission, and $\Delta \eta$ parametrizes the finite longitudinal
extension of the source. $T(x)$ is the freeze-out temperature; if you
don't like the idea of thermalization in heavy ion collisions, you can
think of it as a parameter that describes the random distribution of
the particle momenta at each space-time point around their average
value. The latter is parametrized by a collective flow velocity
$u^\mu(x)$ in the form 
 \begin{equation}
 \label{26}
   u^\mu(x) = \left( \cosh \eta \cosh \eta_t(r), \,
                     \sinh \eta_t(r)\, \bbox{e}_r,  \,
                     \sinh \eta \cosh \eta_t(r) \right) ,
 \end{equation}
with a boost-invariant longitudinal flow rapidity $\eta_l = \eta$ 
($v_l = z/t$) and a linear transverse flow rapidity profile 
 \begin{equation}
 \label{27}
  \eta_t(r) = \eta_f \left( {r \over R} \right)\, .
 \end{equation} 
$\eta_f$ scales the strength of the transverse flow. The exponent of 
the Boltzmann factor in (\ref{3.15}) can then be written as
 \begin{equation}
 \label{26a}
  K\cdot u(x) = M_\perp \cosh(Y-\eta) \cosh\eta_t(r) - 
                \bbox{K}_\perp{\cdot}\bbox{e}_r \sinh\eta_t(r)\, .
 \end{equation}
For vanishing transverse flow ($\eta_f=0$) the source depends only 
on $M_\perp$, and remains azimuthally symmetric for all $K_\perp$.                                                 
Since in the absence of transverse flow the $\beta$-dependent terms in 
(\ref{20b}) and (\ref{20c}) vanish and the source itself depends only 
on $M_\perp$, all three YKP radius parameters then show perfect 
$M_\perp$-scaling. Plotted as functions of $M_\perp$, they coincide 
for pion and kaon pairs (see Fig.~\ref{F1}, left column). For non-zero
transverse flow (right column) this $M_\perp$-scaling is broken by two
effects: (1) The thermal exponent (\ref{26a}) receives an additional
contribution proportional to $K_\perp = \sqrt{ M_\perp^2 - m^2}$. (2)
The terms which were neglected in the second equalities of
(\ref{20b},\ref{20c}) are non-zero, and they also depend on
$\beta_\perp = K_\perp/E_K$. Both effects induce an explicit rest mass
dependence and destroy the $M_\perp$-scaling of the YKP size parameters.  

\section{$\bbox{K}$-dependence of YKP parameters and collective flow}
\label{sec5}

Collective expansion induces correlations between coordinates and 
momenta in the source, and these result in a dependence of the HBT 
parameters on the pair momentum $K$. At each point in the source the 
local velocity distribution is centered around the average fluid 
velocity; two points whose fluid elements move rapidly relative to 
each other are thus unlikely to contribute particles with small 
relative momenta. Essentially only such regions in the source 
contribute to the correlation function whose fluid elements move with 
velocities close to the velocity of the observed particle pair. 

\subsection{The Yano-Koonin velocity and longitudinal flow}
\label{sec5.1}

\begin{figure}
\vspace*{5cm}
\includegraphics{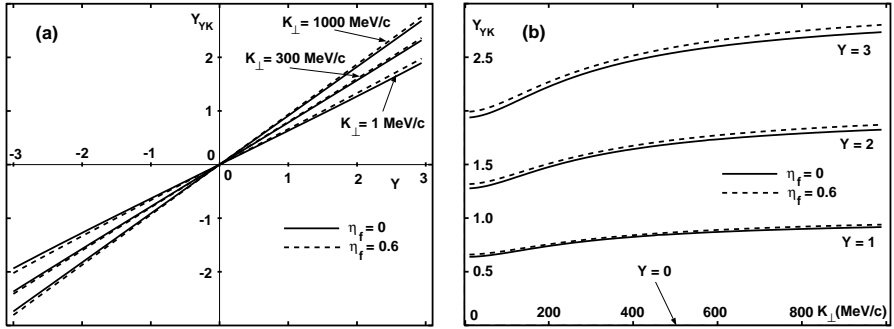}
\caption{(a) The Yano-Koonin rapidity for pion pairs, as a function of
  the pair c.m. rapidity $Y$, for various values of $K_\perp$ and two
  values for the transverse flow $\eta_f$. (b) The same, but plotted
  against $K_\perp$ for various values of $Y$ and $\eta_f$. Source
  parameters as in Fig.~\protect\ref{F1}.
\label{F2}} 
\end{figure}

Fig.~\ref{F2} shows (for pion pairs) the dependence of the YK velocity
on the pair momentum $\bbox{K}$. In Fig.~\ref{F2}a we show the YK
rapidity $Y_{_{\rm YK}} = \frac 12 \ln[(1+v)/(1-v)]$ as a function of
the pair rapidity $Y$ (both relative to the CMS) for different values
of $K_\perp$, in Fig.~\ref{F2}b the same quantity as a function of
$K_\perp$ for different $Y$.  Solid lines are without transverse flow,
dashed lines are for $\eta_f=0.6$.  For large $K_\perp$ pairs, the YK
rest frame approaches the LCMS (which moves with the pair rapidity
$Y$); in this limit all pairs are thus emitted from a small region in
the source which moves with the same longitudinal velocity as the
pair. For small $K_\perp$ the YK frame is considerably slower than the
LCMS; this is due to the thermal smearing of the particle velocities
in our source around the local fluid velocity
$u^\mu(x)$.\cite{HTWW96} The linear relationship between the rapidity
$Y_{_{\rm YK}}$ of the Yano-Koonin frame and the pion pair rapidity
$Y$ is a direct reflection of the boost-invariant longitudinal
expansion flow.\cite{HTWW96} For a non-expanding source $Y_{_{\rm
    YK}}$ would be independent of $Y$. Additional transverse  
flow is seen to have nearly no effect. The dependence of the YK 
velocity on the pair rapidity thus measures directly the longitudinal 
expansion of the source and cleanly separates it from its transverse 
dynamics. 

The NA49 data for 160 A GeV Pb+Pb collisions~\cite{NA49,Sch96} show
very clearly such a more or less linear rise of the Yano-Koonin source
rapidity with the rapidity of the pion pair. This confirms, in the
most transparent way imaginable, their earlier
conclusion~\cite{alber95} based on the $Y$-dependence of the
longitudinal radius pa\-ra\-me\-ter $R_l$ in the Pratt-Bertsch
parametrization that the source created in 200 A GeV S+$A$ collisions
expands longitudinally in a nearly boost-invariant way.

It should be noted that this longitudinal flow need not be of
hydrodynamical (pressure generated) nature. In a description of 
nuclear collisions as a set of longitudinally oriented and
independently fragmenting nucleon-nucleon strings, the string
fragmentation process would also lead to a strong correlation between
the longitudinal positions and momenta of the created hadrons. Thus a
similar linear rise of the YK-rapidity with the pair rapidity would be
expected in jet fragmentation (with the $z$-axis oriented along the
jet axis). It would be interesting to confirm this prediction in
$e^+e^-$ or $pp$ collisions.

\subsection{$M_\perp$-dependence of YKP radii; transverse flow}
\label{sec5.2}

If the source expands rapidly and features large velocity gradients, the
``regions of homogeneity'' contributing to the correlation function
will be small. Their size will be inversely related to the velocity
gradients, scaled by a ``thermal smearing factor'' $\sqrt{T/M_\perp}$
which characterizes the width of the Boltzmann distribution.\cite{CSH95}
If one evaluates the expectation values (\ref{20a}-\ref{20c}) by
saddle point integration one finds for pairs with $Y=0$
 \begin{eqnarray}
 \label{Rs}
    R_\perp^2 &=& R_*^2 \, ,
 \\
 \label{R0}
    R_0^2 &=& (\Delta t_*)^2\, ,
 \\
 \label{Rl}
    R_\parallel^2 &=& L_*^2 \, ,
 \end{eqnarray}
with
 \begin{eqnarray}
 \label{Rstar}
   {1\over R_*^2} &=& {1\over R^2} + {1\over R_{\rm flow}^2}\, , 
 \\
 \label{tstar}
   (\Delta t_*)^2 &=& (\Delta\tau)^2 + 
   2 \left( \sqrt{\tau_0^2 + L_*^2} - \tau_0 \right)^2 \, , 
 \\
 \label{Lstar}
   {1\over L_*^2} &=& {1\over (\tau_0\Delta\eta)^2} 
   + {1\over L_{\rm flow}^2}\, , 
 \end{eqnarray}
where $R_{\rm flow}$ and $L_{\rm flow}$ are the transverse and
longitudinal ``dynamical lengths of homogeneity'' due to the expansion
velocity gradients:
 \begin{eqnarray}
 \label{RH}
   R_{\rm flow}(M_\perp) &=& {R\over \eta_f}\, \sqrt{{T\over M_\perp}}
   = {1\over \partial \eta_t(r)/\partial r} \, \sqrt{{T\over M_\perp}}\, ,
 \\
 \label{LH}
   L_{\rm flow}(M_\perp) &=& \tau_0\, \sqrt{{T\over M_\perp}}
   = {1\over \partial{\cdot}u_l} \, \sqrt{{T\over M_\perp}}\, ,
 \end{eqnarray}
where $u_l$ is the longitudinal 4-velocity.

Thus, for expanding sources, the HBT radius parameters are generically 
decreasing functions of the transverse pair mass $M_\perp$. The slope 
of this decrease grows with the expansion rate~\cite{WSH96,HTWW96}
(this cannot be seen in the saddle point approximated expressions
above). Longitudinal expansion affects mostly the longitudinal radius
parameter $R_\parallel$ and the temporal parameter
$R_0$;\cite{HTWW96} the latter is a secondary effect since particles
from different points are usually emitted at different times, and a 
decreasing longitudinal homogeneity length thus also leads to a 
reduced effective duration of particle emission (see lower panels in
Fig.~\ref{F1}). The transverse radius parameter $R_\perp$ is invariant
under longitudinal boosts and thus not affected at all by longitudinal
expansion (upper left panel in Fig.~\ref{F1}). It begins to drop  
as a function of $M_\perp$, however, if the source expands in the 
transverse directions (upper right panel). Comparing the lower two
left and right panels in Fig.~\ref{F1} one sees that the sensitivity
of $R_\parallel$ and $R_0$ to transverse flow is much
weaker.\cite{HTWW96} Transverse (longitudinal) flow thus affects
mostly the transverse (longitudinal) regions of homogeneity.

While longitudinal ``flow'' is not necessarily a signature for nuclear
collectivity but could be ``faked'' as discussed at the end of the
previous subsection, transverse flow is much more generic in this
respect: there is clearly no transverse collective dynamics in the
ingoing channel, and the only mechanism imaginable for the creation of
transverse flow is multiple (re-)scattering among the participants and
secondaries, leading ultimately to hydrodynamic transverse expansion.

Unfortunately, the observation of an $M_\perp$-dependence of $R_\perp$
by itself is not sufficient to prove the existence of radial transverse 
flow. It can also be created by other types of transverse gradients,
e.g. a transverse temperature gradient.\cite{CSH95,CL96,TH97} To
exclude such a possibility one must check the $M_\perp$-scaling of the
YKP radii, i.e. the independence of the functions $R_i(M_\perp)$ 
($i=\perp,\parallel,0$) of the particle rest mass (which is not broken 
by temperature gradients). Since different particle species are 
affected differently by resonance decays, such a check further 
requires the elimination of resonance effects.  

\section{Resonance decays}
\label{sec6}

Resonance decays contribute additional pions at low $M_\perp$; these 
pions originate from a larger region than the direct ones, due to 
resonance propagation before decay. They cause an 
$M_\perp$-dependent modification of the HBT radii.  

Quantitative studies~\cite{Schlei,WH96} have shown that the resonances 
can be subdivided into three classes with different characteristic 
effects on the correlator: \\
(i) Short-lived resonances with lifetimes up to a few fm/$c$ do not 
propagate far outside the region of thermal emission and thus affect 
$R_\perp$ only marginally. They contribute to $R_0$ and $R_\parallel$
up to about 1 fm via their lifetime; $R_\parallel$ is larger if pion 
emission occurs later because for approximately boost-invariant expansion 
the longitudinal velocity gradient decreases as a function of time.\\ 
(ii) Long-lived resonances with lifetimes of more than several hundred fm/$c$ 
do not contribute to the measurable correlation and thus only reduce the
correlation strength (the intercept at $q=0$), without changing the shape 
of the correlator.\footnote{A reduced correlation strength in the
  two-particle sector could also arise from partial phase coherence in
  the source.\cite{APW93} By comparing two- and three-particle
  correlations, the intercept reducing effects of resonances can be
  eliminated, and the degree of coherence resp. chaoticity in the
  source can be unambiguously determined.\cite{HZ97}}
The reason is that they propagate very far before decaying, thus
simulating a very large source which contributes to the correlation
signal only for unmeasurably small relative momenta.\\ 
(iii) There is only one resonance which does not fall in either of these 
two classes and can thus distort the form of the correlation 
function: the $\omega$ with its lifetime of 23.4 fm/$c$. It contributes
a second bump at small $q$ to the correlator, giving it a non-Gaussian 
shape and complicating~\cite{WH96} the extraction of HBT radii by a 
Gaussian fit to the correlation function. At small $M_\perp$ up 
to 10\% of the pions can come from $\omega$ decays, and this fraction
doubles effectively in the correlator since the other pion can 
be a direct one; thus the effect is not always negligible.  

\begin{figure}
\vspace*{5cm}
\includegraphics{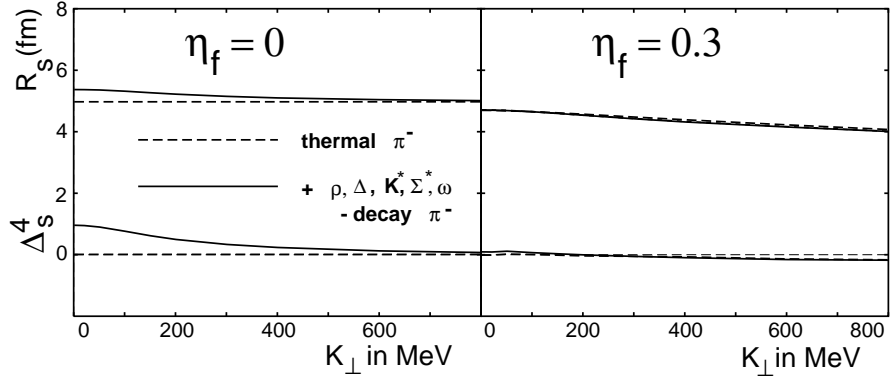}
\caption{The inverted $q$-variance $R_\perp$ and the kurtosis
  $\Delta_\perp$ (the index $s$ in the figure stands for ``sideward'')
  at $Y=0$ as functions of $K_\perp$. Left: $\eta_f=0$ (no transverse
  flow). Right: $\eta_f=0.3$. The difference between dashed and solid
  lines is entirely dominated by $\omega$ decays. Source parameters as
  in Fig.~\protect\ref{F1}, except for $R=5$ fm.
\label{F3}}
\end{figure}

In a detailed model study~\cite{WH96} we showed that resonance 
contributions can be identified through the non-Gaussian features in 
the correlator induced by the tails in the emission function resulting 
from resonance decays. To this end one computes the second and fourth 
order $q$-moments of the correlator.\cite{WH96} The second order 
moments define the HBT radii, while the kurtosis (the normalized
fourth order moments) provide a lowest order measure for the 
deviations from a Gaussian shape. We found~\cite{WH96} that, at least 
for the model (\ref{3.15}), a positive kurtosis can always be 
associated with resonance decay contributions (Fig.~\ref{F3}, left
panel). Strong flow also generates a non-zero, but small and
apparently always negative kurtosis (Fig.~\ref{F3}, right panel).
Any $M_\perp$-dependence of $R_\perp$ which is associated with a positive $M_\perp$-dependent kurtosis must therefore be regarded with suspicion; an 
$M_\perp$-dependence of $R_\perp$ with a vanishing or negative 
kurtosis, however, cannot be blamed on resonance decays.  

In our model, the first situation is realized for a source without 
transverse expansion (left panel of Fig.~1): At small $M_\perp$ 
the $\omega$ contribution increases $R_\perp$ by up to 0.5 fm while 
for $M_\perp > 600$ MeV it dies out. The effect on $R_\perp$ is small 
because the heavy $\omega$ moves slowly and doesn't travel very far 
before decaying. The resonance contribution is clearly visible in the 
positive kurtosis (lower curve). For non-zero transverse flow (right 
panel) there is no resonance contribution to $R_\perp$; this is 
because for finite flow the effective source size for the heavier 
$\omega$ is smaller than for the direct pions, and the $\omega$-decay 
pions thus always remain buried under the much more abundant direct 
ones. Correspondingly the kurtosis essentially vanishes; in fact, it 
is slightly negative, due to the weak non-Gaussian features induced by 
the transverse flow.  

\section{Opaque sources}
\label{sec7}

The emission function (\ref{3.15}) is only one of an infinity of
possible source pa\-ra\-me\-tri\-za\-tions. It is chosen in such a way
that it allows easy implementation of most features which we believe are
important for the sources created in heavy ion collisions. There is,
however, one important physical situation which cannot be parametrized
in any reasonable way by the formula (\ref{3.15}): if the source emits
particles not from the entire volume, but only from a thin surface
layer. This is how the sun radiates photons, and this is also an often
suggested picture for the case that a QGP is created in the collision
which then hadronizes slowly in a deflagration-type strong first order
transition by surface emission of hadrons from the edge of the QGP
blob. 

The significance of such a phenomenon for HBT interferometry was
realized by Heiselberg and Vischer~\cite{HV96} who pointed out that
an effective emission region which is part of a thin surface layer has
a much smaller extension in the ``outward'' or $x$-direction than in
the ``sideward'' or $y$-direction. In other words, such ``opaque
sources'' have $\langle\tilde x^2 - \tilde y^2\rangle < 0$. Depending
on the degree of opacity (the thickness of the surface layer relative
to the source radius) this difference can be large and negative.
The authors pointed out~\cite{HV96} that this leads to the possibility
of a smaller ``outward'' than ``sideward'' HBT radius parameter in the
Pratt-Bertsch parametrization, even at $K_\perp = 0$. Recently
B. Tom\'a\v sik showed~\cite{TH97} that in the YKP parametrization
opacity effects would show up even more spectacularly by a ``lifetime
parameter'' $R_0^2$ which would diverge to $-\infty$ in the limit
$K_\perp \to 0$ resp. $\beta_\perp \to 0$ (see Eq.~(\ref{20c})).

The source (\ref{3.15}) can be made opaque by multiplying it by the
factor~\cite{HV96,TH97}
 \begin{equation}
 \label{opaque}
   \exp\left( -\kappa {l_{\rm eff} \over \lambda} \right)
 \end{equation}
where $\lambda$ is the mean free path, 
\begin{equation}
  l_{\rm eff} = l_{\rm eff}(r,\phi) = 
  e^{- \frac{y^2}{2R^2}} \, \int_x^{\infty} \,
  e^{- \frac{{x'}^2}{2R^2}} \, dx' \qquad {\rm with  } \quad
  y = r\, \sin \phi ,\: x = r\, \cos\phi\, .
\label{op3.7}
\end{equation}
is the effective travelling distance of the emitted particle through
matter in the source (\ref{3.15}), and $\kappa=\sqrt{8/\pi}$ is a
parameter which is adjusted in such a way that particles emitted from
the center suffer the same suppression as in the model of Heiselberg
and Vischer~\cite{HV96} who use a box distribution instead of the
Gaussians in (\ref{3.15}).

\begin{figure}
\vspace*{5cm}
\includegraphics{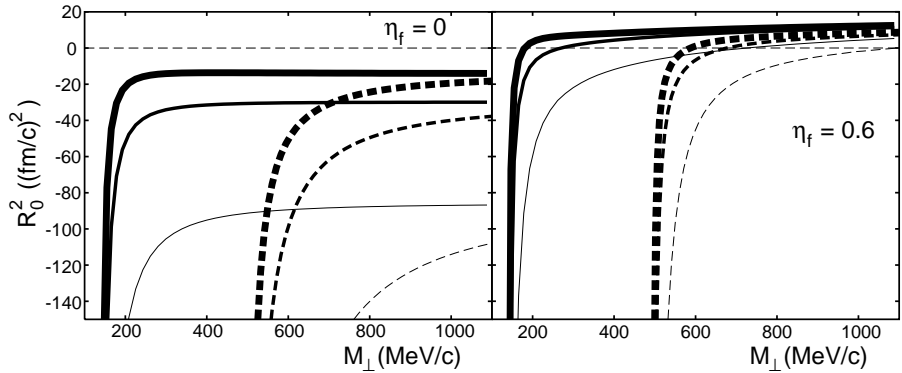}
\caption{The YKP parameter $R_0^2(M_\perp)$ for midrapidity pairs for 
   various combinations of $\lambda$ and $\eta_f$. Solid (dashed) 
   lines are for pions (kaons). Thin lines: $\lambda = 1 {\rm \ fm} = 
   R/7$; medium lines: $\lambda = 7 {\rm \ fm} = R$; thick lines: 
   $\lambda = 14 {\rm \ fm} = 2R$. Left: $\eta_f = 0$; right: $\eta_f
   = 0.6$. Source parameters: $T=100$ MeV, $\Delta\eta =1.3$,
   $\tau_0=7.8$ fm/$c$, $\Delta_\tau = 2$ fm/$c$, $R=7$ fm. 
\label{F4}}
\end{figure}

Fig.~\ref{F4} shows the interesting YKP parameter, the ``temporal''
radius parameter $R_0^2$, as a function of $M_\perp$ for sources with
different degrees of opacity. With or without transverse flow, the
crucial features of opacity are clearly visible: the negative
contribution $\sim \langle\tilde x^2 - \tilde y^2\rangle$ in
(\ref{20c}) drives $R_0^2$ to negative values at small $K_\perp$, and
this happens the sooner the shorter the mean free path $\lambda$,
i.e. the thinner the surface layer is.

Comparing the solid curves for pion pairs in Fig.~\ref{F4} with the
Pb+Pb data~\cite{Sch96} in Fig.~\ref{F5} below I conclude that mean
free path values $\lambda < R$ are essentially excluded. (This
conclusion has in the meantime been checked by more extensive
parameter studies.) In other words, the sources created in 160 A GeV
Pb+Pb collisions are not at all opaque, but rather ``transparent'',
meaning that particles are emitted from the whole volume by bulk
rather than surface dominated freeze-out.

\section{Analysis of Pb+Pb data}
\label{sec8}

\begin{figure}
\vspace*{15cm}
\includegraphics{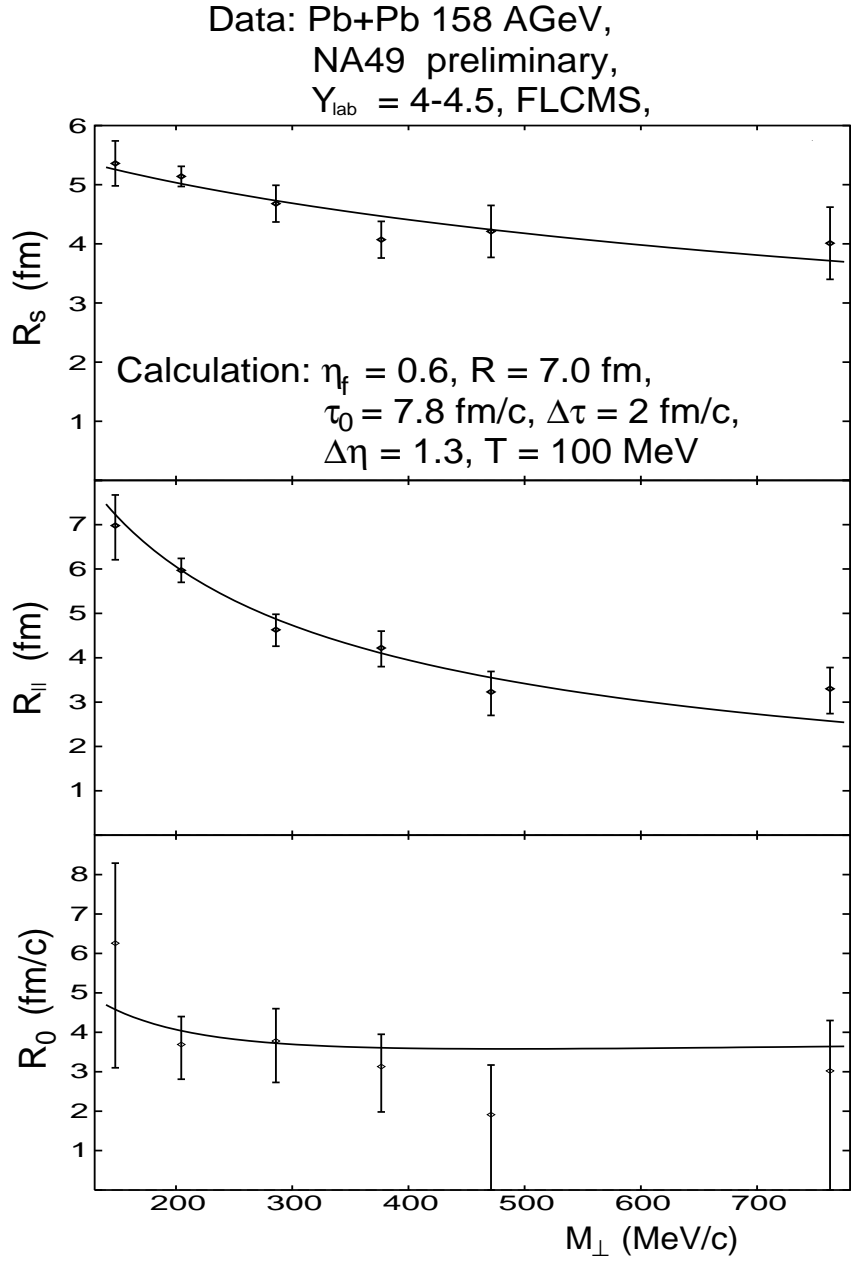}
\caption{ 
 $R_\perp$, $R_\parallel$ and $R_0$ for 158 A GeV/$c$ Pb+Pb collisions 
 as functions of the transverse pair momentum. The data are from the 
 NA49 Collaboration.\protect\cite{Sch96} The lines are a fit with the model 
 (\protect\ref{3.15}), with fit parameters as given in the figure.  
 \label{F5}}
\end{figure}

In Fig.~\ref{F5} we show a numerical fit of the YKP radius parameters,
using the expressions (\ref{20a})-(\ref{20c}) with our model source 
(\ref{3.15}), to data collected by the NA49 collaboration in 158 A 
GeV/$c$ Pb+Pb collisions.\cite{NA49,Sch96} Please note that this fit
refers to only a single rapidity slice of the available data, and it 
does not include resonance decays (although we do not expect the
latter to change things much, except for reducing $\tau_0$ by about 1
fm/$c$, the average lifetime of the shortlived resonances, see
Sec.~\ref{sec6}). The fit result must therefore be taken with some
care. A comprehensive simultaneous analysis of all single particle
spectra and two-particle correlation data from Pb+Pb collisions is in
progress.   

After $\Delta \eta{=}1.3$ has been adjusted to reproduce the 
width of the pion rapidity distribution,\cite{Sch96} the parameters 
$\tau_0$ and $\Delta \eta$ are essentially fixed by the magnitude of 
$R_\parallel$ and $R_0$. The radius $R$ is fixed by the magnitude of 
$R_\perp(K_\perp=0)$ once the temperature $T$ and transverse flow 
$\eta_f$ are known. The $M_\perp$-dependence of $R_\perp$ fixes $T$ 
and $\eta_f$, albeit not independently: essentially only the 
combination $\eta_f \sqrt{M_\perp/T}$, i.e. the velocity gradient 
divided by the thermal smearing factor, can be
extracted.\cite{CNH95,Sch96} This is similar to the single particle
spectra whose $M_\perp$-slopes determine only an effective blushifted 
temperature,\cite{LH89} $T_{\rm eff} = T \sqrt{{1+ \bar v_f \over 1-
    \bar v_f}}$. The correlations between $T$ and $\eta_f$ are,
however, exactly opposite in the two cases: for a fixed spectral slope
$T$ must be decreased if $\eta_f$ increases while a fixed
$M_\perp$-slope of $R_\perp$ requires decreasing values of $\eta_f$ if
$T$ is reduced.\cite{Sch96} {\em The combination of single-particle
  spectra and two-particle correlation thus allows for a separate
  determination of $T$ and $\eta_f$.} 

For the fit in Fig.~\ref{F5} the freeze-out temperature was set by hand
to $T=100$ MeV. The resulting flow parameter $\eta_f{=}0.6$
corresponds to an average transverse flow velocity $\bar v_f{=}0.58$.
This combination of $T$ and $\eta_f$ results in single-particle 
spectra with roughly the right shape. Somewhat higher temperature
values of around $T=120$ MeV as advocated by
K\"ampfer~\cite{kaempfer96} with an average transverse flow velocity
$\bar v_f = 0.43\,c$ produce a somewhat flatter decrease of $R_\perp$
with $M_\perp$ but still appear to be consistent with the data inside
the error bars. Fitting $R_\perp$ with even higher temperatures
results in larger $\eta_f$-values which leads to single particle
spectra which are much too flat. 

Let us discuss in more detail the numbers resulting from this fit. 
First, the transverse size parameter $R{=}7$ fm is surprisingly large. 
Resonance contributions are not expected to reduce it by more than 0.5 
fm.\cite{WH96} The transverse flow correction to $R_\perp$ is
appreciable, resulting in a visible transverse homogeneity length of  
only about 5.5 fm at small $K_\perp$, but even this number is large. 
$R{=}7$ fm corresponds to an r.m.s. radius $r_{\rm rms} = 
\sqrt{\langle \tilde x^2 + \tilde y^2 \rangle} \approx 10$ fm of the 
pion source, to be compared with an r.m.s. radius $r_{\rm rms}^{\rm 
Pb} = 1.2 \times A^{1/3} * \sqrt{2/5}$ fm = 4.5 fm for the density 
distribution of the original Pb nucleus projected on the transverse 
plane.\footnote{In my lecture given at the workshop I made an
  embarrassing error by a factor $\sqrt{2}$ which is also contained in
  the writeup~\cite{hirschegg} of my talk at Hirschegg in January
  1996. I would like to thank D. Ferenc for pointing out this error.}
This implies a transverse expansion of the reaction zone by a linear
factor 10/4.5 = 2.2. That we also find a large transverse flow velocity 
renders the picture consistent. The longitudinal size of the collision 
region at the point where the pressure in the system began to drive 
the transverse expansion can be estimated as follows: for the source
to expand in, say, the $y$-direction from $\sqrt{\langle y^2
  \rangle}_{\rm initial}= 1.2\, A^{1/3} /\sqrt{5}$ fm = 3.2 fm to
$\sqrt{\langle y^2 \rangle}_{\rm final} = R = 7$\,fm with an average
transverse flow velocity of at most $\bar v = 0.58\, c$ (the
freeze-out value determined from the fit of $R_\perp$) requires at
time of at least $(7-3.2)/0.58$ fm/$c$ = 6.5 fm/$c$. Due to the
selfsimilarity of the longitudinal expansion the longitudinal
dimension of the source grows linearly with $\tau$. If the total
expansion time until freeze-out is given by the fit parameter $\tau_0
= 7.8$ fm/$c$, the source expanded in the 6.5 fm/$c$ during which
there was transverse expansion by a factor 7.8/(7.8-6.5) = 7.8/1.3 = 6
in the longitudinal direction. We conclude that the fireball volume
must have expanded by a factor $6 * 2.2^2 \approx 30$ between the
onset of transverse expansion and freeze-out! This is the clearest
evidence for strong collective dynamical behaviour in
ultra-relativistic heavy-ion collisions so far.   

The local comoving energy density at freeze-out can be estimated from 
the fitted values for $T$ and $\eta_f$. The thermal energy density of 
a hadron resonance gas at $T=100$ MeV and moderate baryon chemical 
potential is of the order of 50 MeV/fm$^3$. The large average 
transverse flow velocity of $\langle v_f \rangle \approx 0.58$ implies 
that about 50\% flow energy must be added in the lab frame. This 
results in an estimate of about $0.050$ GeV/fm$^3 \times 1.5 \times 30
\approx$ {\bf 2.2 GeV/fm}$^3$ for the energy density of the reaction zone at 
the onset of transverse expansion.  This is well above the critical 
energy density $\epsilon_{\rm cr} \leq 0.9$ GeV/fm$^3$ predicted by 
lattice QCD for deconfined quark-gluon matter.\cite{K96} Whether this 
energy density was fully thermalized is, of course, a different
question. It must, however, have been accompanied by transverse
pressure (i.e. some degree of equilibration of momenta must have
occurred already before this point), because otherwise transverse
expansion could not have been initiated.   
                                        
\section{Conclusions}
\label{sec9}

I hope to have shown that

\begin{itemize}

\item
 two-particle correlation functions from heavy-ion collisions provide 
valuable information both on the geometry {\bf and} the dynamical 
state of the reaction zone at freeze-out;

\item
 a comprehensive and simultaneous analysis of single-particle spectra 
and two-particle correlations, with the help of models which provide
a realistic parametrization of the emission function, allows for an 
essentially complete reconstruction of the final state of the reaction 
zone, which can serve as a reliable basis for theoretical 
back-extrapolations towards the interesting hot and dense early stages 
of the collision;
 
\item
 simple and conservative estimates, based on the crucial new 
information from HBT measurements on the large transverse size of the 
source at freeze-out and using only energy conservation, lead to the 
conclusion that in Pb+Pb collisions at CERN, before the onset of 
transverse expansion, the energy density exceeded comfortably the 
critical value for the formation of a color deconfined state of quarks 
and gluons. There is, however, no evidence for long time delays due to
hadronization of the QGP, and pion freeze-out occurs in bulk rather
than from the surface of the collision fireball. This is in line with
lattice results which predict at most a {\em weakly} first order
confinement transition, and with other evidence~\cite{LTHSR95} for
rapid hadronization.

\end{itemize}

\section*{Acknowledgements:} 
 This work was supported by grants from DAAD, DFG, NSFC, BMBF and GSI.
 The results reported here were obtained in collaboration with
 D. Anchishkin, S. Chapman, P. Scotto, B. Tom\'a\v sik, U. Wiedemann,
 and Y.-F. Wu, to whom I would like express my thanks. I gratefully
 acknowledge many discussions with H. Appelsh\"auser, T. Cs\"org\H o,
 D. Ferenc,  M. Ga\'zdzicki, S. Sch\"onfelder, P. Seyboth, and
 A. Vischer. 


\end{document}